\documentclass[tighten]{emulateapj}

\slugcomment{Preprint, \today}

\shorttitle{Gemini Laser-AO on 4C+41.17}
\shortauthors{Steinbring}

\input epsf
\def\plotone#1{\centering \leavevmode
\epsfxsize=1.00\columnwidth \epsfbox{#1}}

\def\plotonewide#1{\centering \leavevmode
\epsfxsize=2.05\columnwidth \epsfbox{#1}}
\def\plotonenarrow#1{\centering \leavevmode
\epsfxsize=0.70\columnwidth \epsfbox{#1}}

\begin{document}

\title{A Star-Forming Shock Front in Radio Galaxy 4C+41.17 \\ Resolved with Laser-Assisted Adaptive Optics Spectroscopy}

\author{Eric Steinbring\altaffilmark{1}}

\altaffiltext{1}{National Research Council Canada, Victoria, BC V9E 2E7, Canada; Eric.Steinbring@nrc-cnrc.gc.ca}

\begin{abstract}

Near-infrared integral-field spectroscopy of redshifted [O {\small III}], H$\beta$ and optical continuum emission from $z=3.8$ radio galaxy 4C+41.17 is presented, obtained with the laser-guide-star adaptive optics facility on the Gemini North telescope. Employing a specialized dithering technique, a spatial resolution of 0\farcs10 or 0.7 kpc is achieved in each spectral element, with velocity resolution of $\sim70~{\rm km}~{\rm s}^{-1}$. Spectra similar to local starbursts are found for bright knots coincident in archival {\it Hubble Space Telescope} (HST) restframe-ultraviolet images, which also allows a key line diagnostic to be mapped together with new kinematic information. There emerges a clearer picture of the nebular emission associated with the jet in 8.3 GHz and 15 GHz Very Large Array maps, closely tied to a Ly$\alpha$-bright shell-shaped structure seen with HST. This supports a previous interpretion of that arc tracing a bow shock, inducing $\sim10^{10-11}~M_\odot$ star-formation regions that comprise the clumpy broadband optical/ultraviolet morphology near the core.

\end{abstract}


\section{Introduction}\label{introduction}

Radio galaxies, by virtue of their intrinsic brightnesses, provide a ready means of probing star formation back to early times. The strong correlation of their near-infrared (NIR) $K$-band emission with redshift - the ``$K-z$" relation - is consistent with a brief epoch of rapid star formation ($z\sim 5 - 10$) followed by passive evolution \citep{Lilly1984, Eales1993, Eales1997, Inskip2002, Willott2003}. For the most powerful high-redshift radio galaxies (HzRGs) with $z>1$, the orientation of nonthermal radio output coincides with the restframe ultraviolet (UV) galaxy, producing a well-known ``alignment effect" \citep{McCarthy1987, Chambers1987} which may be primarily the result of interaction between the jet and intergalactic gas inducing star formation, but could in general include some contribution from nebular spectral lines and scattered active galactic nucleus (AGN) light by either electrons or dust.

The ultrasteep spectrum HzRG 4C+41.17 ($z=3.787$, 36 Jy at 26 MHz, 2.7 Jy at 178 MHz, $\alpha=-1.4$; Chambers et al., 1990) is of particular interest because its clumpy asymmetric optical morphology and spectrum similar to local starbursts - combined with low polarization - provides direct evidence for jet-induced shocks forming stars \citep[e.g.][]{Dey1997, Bicknell2000, Reuland2007, Rocca-Volmerange2013}. This could also be important in the context of AGN feedback due to its massiveness, which at $K=19$ mag \citep{Steinbring2002} is approaching the $10^{12}~M_\odot$ limit of RGs \citep{Rocca-Volmerange2004} despite a 12 Gyr look-back time. A cosmology with $\Omega_\Lambda = 0.73$, $\Omega_{\rm m}=0.27$ and $H_0=71~{\rm km}~{\rm s}^{-1}~{\rm Mpc}^{-1}$ is maintained throughout.

\begin{figure*}
\plotonewide{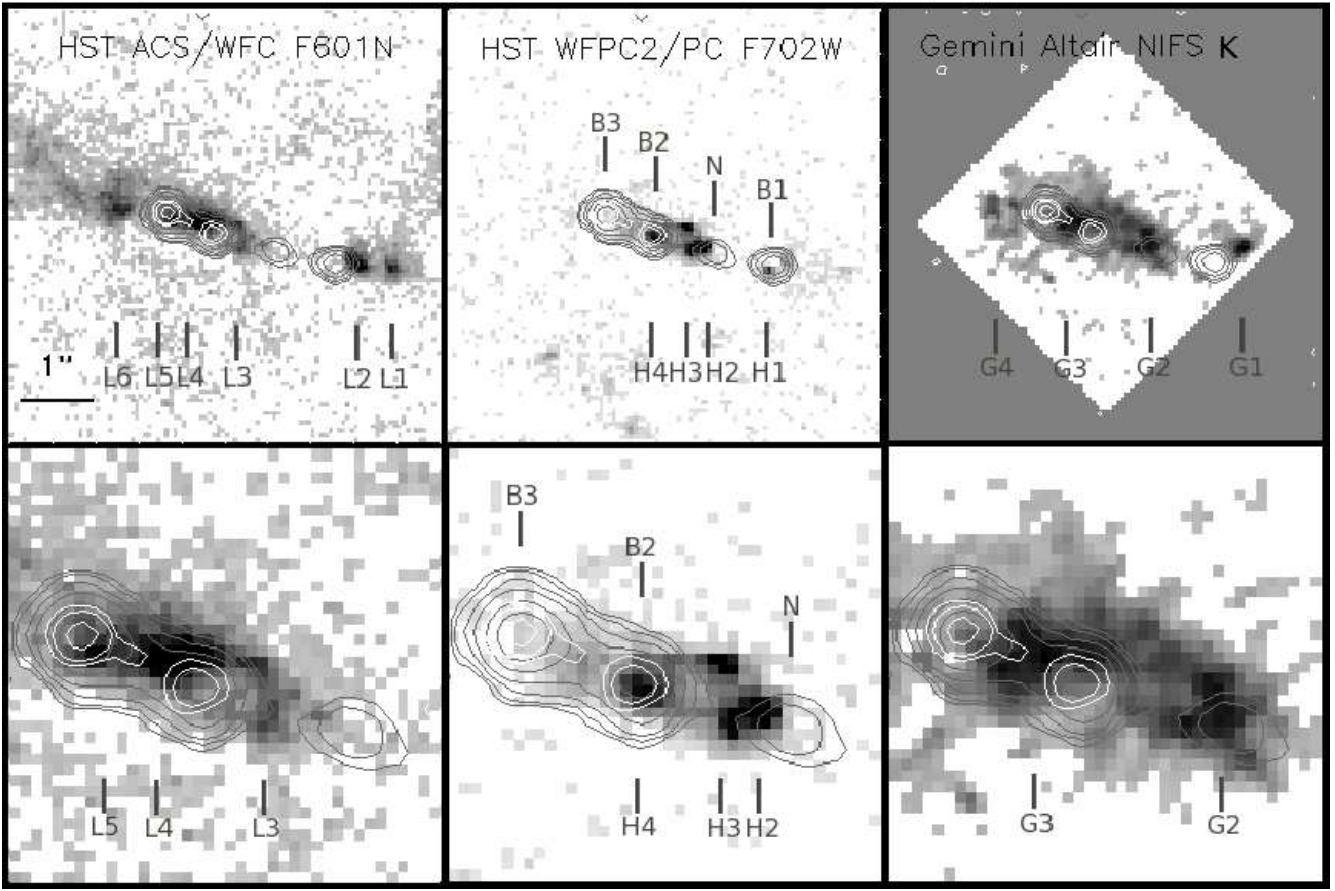}
\caption{Archival WFPC2/PC F702W image of 4C+41.17 (middle column of two panels) superimposed with available VLA A-array 8.3 GHz (grey contours) and 15 GHz (white) radio maps. Also shown: HST ACS/WFC FR601N (left panels); Gemini ALTAIR broadband $K$ (right) produced from the new observations discussed here by collapsing the NIFS datacube in the spectral dimension. Datacube processing is discussed more in Section~\ref{analysis}. Bottom panels are central $2\arcsec\times2\arcsec$ sections of each image. North is up and east is left and the same logarithmic scaling is used throughout.}
\label{plot_images}
\end{figure*}

The orientation of extended Ly$\alpha$ emission in 4C+41.17, over 15\arcsec~across or 110 kpc (7.23 kpc ${\rm arcsec}^{-1}$) lies tightly along the axis of the two main radio components, `A' and `B' in the original notation of Chambers et al. This correlation continues down to sub-0\farcs5 scales within B at 8.3 GHz and 15 GHz in Very Large Array (VLA) maps, with 0\farcs2 full-width at half-maximum (FWHM) beam sizes \citep{Carilli1994}. {\it Hubble Space Telescope} (HST) Wide-Field and Planetary Camera 2 (WFPC2) F702W filter images further resolved the galaxy into compact knots \citep{Miley1992}, the brightest of which is close to the radio nucleus, `N'. The HST identifications, denoted by `H', are all adopted here, with archival images (and radio contour maps) displayed in Figure~\ref{plot_images} in the middle panels; the central region spans less than 2\arcsec~or 15 kpc. Velocities approaching $2000~{\rm km}~{\rm s}^{-1}$ in this inner part of the galaxy were found from seeing-limited Canada-France-Hawaii Telescope (CFHT) integral field unit (IFU) spectroscopy of Ly$\alpha$ \citep{Adam1997}, which may indicate outflows of gas on this and larger scales. Narrower emission lines, e.g. ${\rm FWHM}\approx500~{\rm km}~{\rm s}^{-1}$ for the ${\rm C~{\scriptstyle IV}}~\lambda\lambda1548, 1550$ doublet (not subject to strong resonant scattering effects like Ly$\alpha$) were found with deep Keck Telescope spectropolarimetry, along with an absorption-line spectrum similar to that of star-forming regions for nearby galaxies, including P-Cygni-like profiles and essentially unpolarized (2-$\sigma$ limit of $P<2.4$\%) rest-frame UV continuum \citep{Dey1997}. At near-solar metallicities and plausible gas densities of $n_{\rm H} = 1-10~{\rm cm}^{-3}$, cooling timescales are sufficiently short ($\sim 10^6~{\rm yr}$) that shocks in this velocity range ($v_{\rm sh}\sim500~{\rm km}~{\rm s}^{-1}$) are fully radiative \citep{Dopita1995, Dopita1996} and \cite{Bicknell2000} proposed that a Ly$\alpha$-bright arc-shaped structure visible in a 0\farcs1-resolution HST WFPC2 FR601 ramp-filter image was the leading edge of an elliptical bow shock, crossing between `B2' and `B3' at its western apex, at position `L4' in the left-hand panels of Figure~\ref{plot_images}. Regions of stars should be close to this point of interaction causing fragmentation and collapse, with precursor emission coming from further along the jet axis towards B3, the radio lobe. Notably, B3 is near another Ly$\alpha$ ``hot spot", `L5', in an HST Advanced Camera for Surveys / Wide Field Camera (ACS/WFC) FR601N ramp-filter image\footnote{Processed HST data obtained from the Canadian Astronomical Data Centre Hubble Legacy Archive.} discussed in \cite{vanBreugel2006} and shown in Figure~\ref{plot_images}; a comparison of previous analysis to these newer archival data will be presented here.

Adaptive optics (AO) NIR broadband imaging with CFHT found an elongated structure with the bulk of the $K$-band light (roughly restframe $B$) between B2 and B3 at its resolution limit of 0\farcs3 \citep{Steinbring2002} and an ${\rm F702W}-K\geq2$ colour consistent with a young age, including possible reddening. This is also approximately the locus of a $\sim6\times10^{10}~M_\odot$ CO system observed with millimetre interferometery \citep{DeBreuck2005}, indicative of 4C+41.17 possessing a significant reservoir of gas. Recent spectral energy distribution (SED) fitting by \cite{Rocca-Volmerange2013} explains this correspondence as two distinct stellar components; one of $10^{11}~M_\odot$ roughly 30 Myr old, from a brief (1 Myr) starburst, and a second massive $10^{11-12}~M_\odot$ evolved population, already $\sim 1~{\rm Gyr}$ old at $z\approx4$. A possibility is that 4C+41.17 might be undergoing assembly, yet still with active star formation, as this would be generally consistent with it being at the centre of a protocluster environment, based on overdensities in {\it Herschel} $70~\mu{\rm m}$ through $500~\mu{\rm m}$ and {\it Spitzer Space Telescope} photometry \citep{Wylezalek2013}.

This paper reports on new AO observations of 4C+41.17 using the Near-infrared Integral-Field Spectrograph (NIFS), an image-slicing IFU, combined with the ALTtitude conjugate Adaptive optics for the InfraRed (ALTAIR) laser facility on Gemini North. An advantage to AO is high sensitivity to compact structures, gaining better contrast for those against the background rather than improving detection of fainter, diffuse emission. The intention was to merge the VLA data with spectroscopy in $K$ band, at better spatial resolution than was obtained for CFHT AO, and look specifically at the spatial correspondance with Ly$\alpha$ emission seen with HST near the galaxy core. These observations are presented in Section~\ref{data}. Evidence for clumpy star-forming regions associated with shocks is presented in Section~\ref{analysis}. Comparison of kinematics and [O {\small III}]$\lambda$5007 to Ly$\alpha$ ratio with updated models of fast radiative shocks is discussed in Section~\ref{discussion} along with implications for the star-formation history of 4C+41.17 and possibly other massive HzRGs.

\section{Observations and Calibration}\label{data}

Gemini North observations occured on the nights of 2008 December 18 and 2009 February 18 with NIFS \citep{McGregor2003}. All data were obtained in the $K$ filter, with the grating set to a central wavelength of 2.25 $\mu$m. NIFS operated behind ALTAIR, which provides AO correction with either a natural guide star (NGS) or that plus a sodium laser beacon. A dichroic allows sensing in light shortward of $1~\mu{\rm m}$, separated from the NIR science path. In laser-guide-star (LGS) mode \citep{Boccas2006} the wavefront sensor (WFS) and deformable mirror for correction of high-order aberrations are both along the telescope (on-axis) optical path to the target, while atmospheric tip and tilt compensation employs a separate flat mirror guided by an NGS nearby on the sky. An $R=14$ mag star at ${\rm R.A}=$~06h50m50.638s ${\rm Dec.}=$~+41$^\circ$30\arcmin49\farcs70, 24\arcsec~away from the target was employed for the latter purpose, the same star used in \cite{Steinbring2002} for NGS mode AO observations on CFHT. In this case, as the target is faint and the NIFS field of view (FoV) is small ($3\arcsec \times 3\arcsec$) offsets were made ``blind" relative to the NGS. The NIFS field was centred at ${\rm R.A}=$~06h50m52.093s ${\rm Dec.}=$~+41$^\circ$30\arcmin30\farcs18 (J2000.0) - the barycenter of emission in HST F702W - and rotated to a position angle of 225 degrees (east of north). This rotation angle also allowed an efficient fit of the square NIFS field of view to the optical/NIR galaxy, approximately $4\arcsec$ long east-west, and simultaneously took better advantage of the non-circular (slightly egg-shaped) patrol region of the ALTAIR tip-tilt WFS - providing an extra 2\arcsec~clearance for offsets along that particular axis. Sky subtraction was obtained by employing a target-sky-target-target-sky-target pattern using offsets of 6\arcsec~to the north, which avoided a faint star between the target and the guide star, as well as a low-$z$ galaxy still further to the north. Individual exposures were 600 s, with each sequence followed by a spectrum of an arc lamp, and bracketed by spectra of bright telluric standard stars. The sky-subtracted target observations were stacked to produce a datacube by registering the frames using the commanded telescope offsets, for a total integration on target of 7200 s.

\subsection{Point-spread Function and Sub-slice Dithers}

Instrument performance was characterized using the standard star observations, which bracketed those of the target. These calibrations were obtained in NGS mode, not LGS. Guiding on bright ($R=8$) stars ensured optimal ALTAIR on-axis AO correction under the good natural seeing ($\leq0\farcs8$ at 0.5 $\mu$m) experienced. Consequently, this procedure provided a solid, regular check on the repeatability of offsets within the telescope plus ALTAIR NIFS system, that is, confirming with near-Nyquist-sampled stellar centroids (at 2.3 $\mu$m) to be within an individual NIFS pixel (0\farcs043) over the course of observations. The small FoV of NIFS did not include a faint star near the target for direct point-spread function (PSF) determination, and as performance of AO (in either NGS or LGS mode) varies strongly with seeing and the vertical distribution of turbulence in the atmosphere - which was not monitored - a reasonable range for the delivered PSF FWHM at the target position is instead estimated.

To first order, the AO PSF can be modelled as a diffraction-limited Gaussian core superimposed on a seeing-limited uncorrected halo. The on-axis PSF FWHM was near 0\farcs073 with an on-axis Strehl ratio $S_0$ (PSF peak relative to that of a perfect diffraction pattern) approaching 0.40, the best-possible with ALTAIR NGS. Under these conditions, the performance of ALTAIR LGS is similar, affected primarily by focus anisoplanatism (``cone effect") and can provide $S\approx0.20$ to 0.25. Degradation due to angular anisoplantism (imperfectly sensed and corrected wavefront tip and tilt, sometimes referred to as anisokinetic error) induced by the NGS to target/laser offset is a weaker effect, including slight PSF elongation along the axis towards the NGS (to the northwest). In this Strehl-ratio regime, the PSF is still dominated by the core rather than halo, and FWHM grows roughly as $\sqrt{S_0/S}$ \citep{Steinbring2002}, so the LGS PSF FWHM of observations could conceivably have been better than $0\farcs090$ ($S/S_0<1.5$) and not likely worse than $0\farcs073\times\sqrt{0.40/0.10}=0\farcs146$.

The choice of dither pattern during target observations ensured preservation of the spatial resolution of the PSF core in the output datacube. For a single pointing the best-achievable spatial resolution with NIFS is limited by the width of a slice. The NIFS IFU divides the $2.987\arcsec\times2.987\arcsec$ field into 29 slices, with each slice being ~0\farcs103 wide. Along each slice the spatial sampling is finer, 0\farcs043 per pixel. In order to optimize spatial resolution across slices the 6 dither positions of the observations were in equal steps spanning 0\farcs200 across slices. This provided over-sampling akin to the effect of ``drizzling" as employed for HST \citep{Fruchter2002}, re-distributing every two slices into 5 equally-wide bins of 0\farcs040. This resulted in effectively the same spatial sampling across and along slices, but was of sufficiently restricted span to avoid inducing cross-talk with a third slice, a problem encountered with previous NIFS spectroscopy \cite{Steinbring2011}. It should be pointed out that as much as one pixel in ``blurring" of the PSF may have been added in this way due to imperfect mechanical repeatability (slop) in the telescope dithers, that is, ${\rm FWHM}=\sqrt{(0\farcs043)^2+(0\farcs090)^2}=0\farcs100$, but this would still have improved on the resolution allowed by the width of a single slice.

\begin{figure*}
\plotonewide{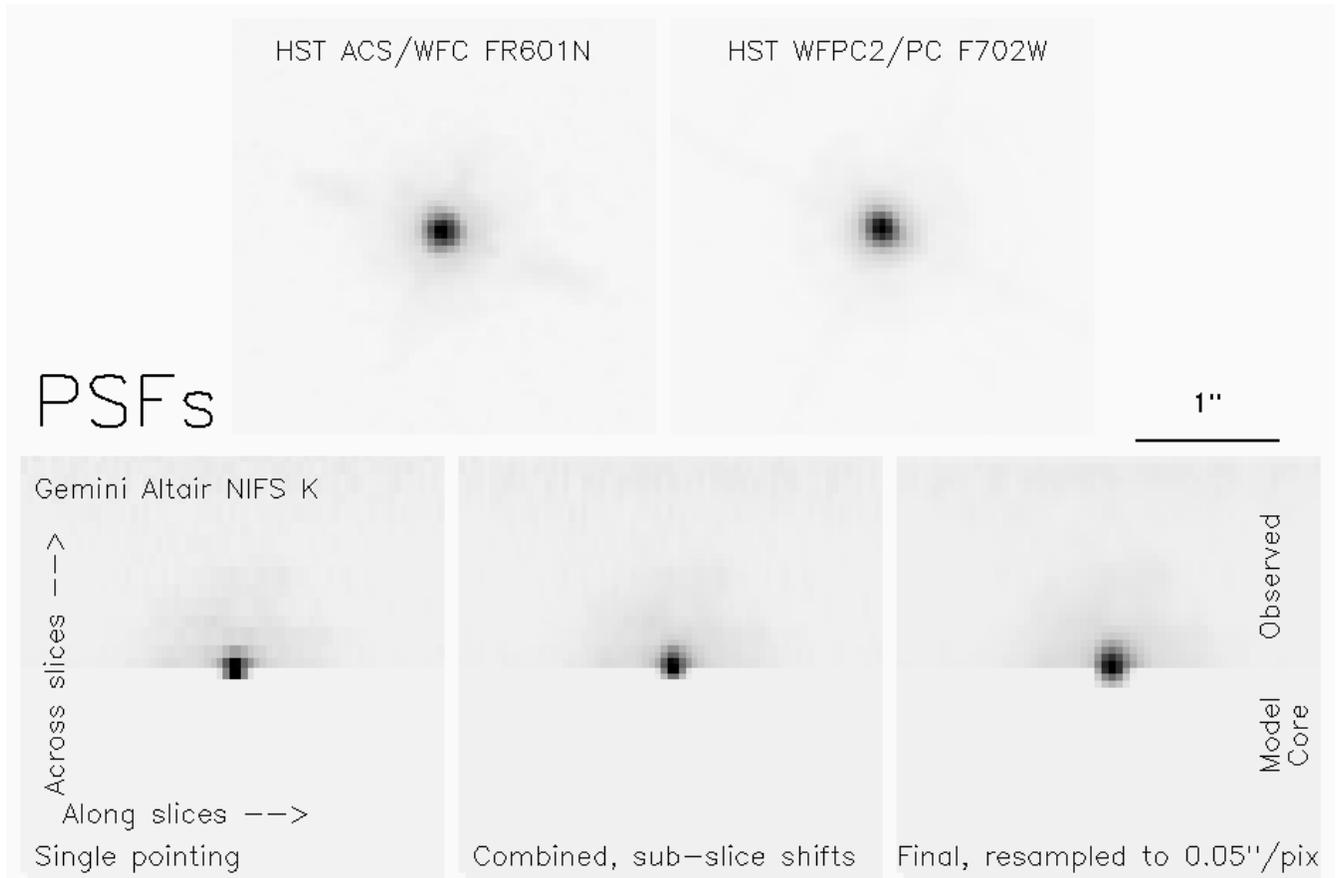}
\caption{PSFs for HST ACS/WFC FR601N and WFPC2/PC F702W (top row) compared to Gemini ALTAIR NIFS $K$ using sub-slice dithering (bottom row). Linear scaling is used throughout, truncated at the half-peak flux. The top half of each NIFS panel shows the result obtained from calibration observations of a star, that with a Gaussian model of the core is shown below. Note how the resampled, dithered NIFS PSF (at far right) is more similar to HST than than for a single pointing (far left).}
\label{plot_psfs}
\end{figure*}

A corresponding NIFS $K$-band PSF is shown in Figure~\ref{plot_psfs}, obtained from applying the same target dither pattern to the standard star observations, a single pointing of which is displayed at far left. As these calibrations were taken with NGS AO correction on a bright star, they represent a best case for the PSF core, without the penalties of anisoplanatism, but including possible positional error. The lower half of the panel shows a model of the PSF core, a Gaussian of $0\farcs100$ along a slice, and with flux distributed evenly across each slice, a top-hat function with a width of 0\farcs103. At far right is the final PSF rebinned to 0\farcs050 pixels. Apart from the seeing-limited halo, the resampled model and observed PSFs agree, and are comparable with HST: the top panels show the ACS/WFC FR601N (0\farcs049~${\rm pix}^{-1}$) WFPC2/PC F702W (0\farcs045~${\rm pix}^{-1}$) PSFs from unsaturated stars in the field.

A side benefit of the chosen dither pattern was that it also facilitated, via standard median-clipping methods, removal of NIFS bad pixels and cosmic ray strikes, the latter of which can be irregular and corrupt up to three contiguous pixels.

\subsection{Wavelength Calibration of Datacube}

Wavelength calibration was obtained from the arc-lamp spectra, with telluric and flux calibration determined from the spectra of the standard stars; best sensitivity in redshifted emission spanned from 4686\AA~ to 5073\AA, and yielded a spectral resolution of $\lambda/\delta\lambda \approx 6360$ with a FWHM velocity resolution of $66~{\rm km}~{\rm s}^{-1}$. A combination of standard IRAF packages and custom Interactive Data Language (IDL) utilities were used. The datacube was re-binned to a pixel scale of $0\farcs050$ without smoothing in the spatial axes. Smoothing was applied only in the spectral dimension, with a Gaussian filter of ${\rm FWHM} = 15$\AA, well matched to the final instrumental resolution.

\section{Reductions and Analysis}\label{analysis}

As a first step, a broadband image approximating $K$ was synthetized by summing the NIFS datacube in each spatial pixel, that is, along its full spectral dimension without excluding emission or sky-lines. This is shown in the right-hand column of Figure~\ref{plot_images}; available HST and VLA data are those described in Section~\ref{introduction}. Fainter emission is more extended than in a previous CFHT AO image \citep{Steinbring2002}, although some of this could be the consequence of deeper imaging, and light scattered into the uncorrected halo of the Gemini AO PSF. 

Four main stuctures visible are indicated by the labels `G1' through `G4' in Figure~\ref{plot_images}. The per-pixel $S/N$ of these structures are over 10, but it falls below 3 in the outer diffuse regions. The orientation of the NIFS field with a 45 degree rotation evidently helped fit the galaxy better, although dithering impared detection near the southwest and northeast edges. The fall-off in the two pixels nearest the verge at knot G1 is probably an artifact of this, rendering any centroid of its position suspect, and it also impacts faint emission in the northeast corner near G4. Sensitivity is uniform in the central 2\arcsec-square region, which is the focus of the analysis that follows. Contamination from the PSF halo was limited by further restricting this to pixels with greater than 50\%~of the peak flux; those with less were found to provide too poor $S/N$ and were excluded from further analysis.

\subsection{Central Galaxy and Astrometry}\label{astrometry}

There is a striking similarity between the clumpy structure in the $K$ image and that seen with HST. Bright knots in the WFPC2/PC F702W and ACS/WFC FR601N Ly$\alpha$ images in Figure~\ref{plot_images} are those discussed in \cite{Chambers1996}, and their astrometry is maintained here, as are coordinates for the VLA 8.3 GHz and 15 GHz maps reported in \cite{Carilli1994}. Relative positional accuracy of the maps can be expected to be within $0\farcs05$ (better at the higher frequency) with the minor resampling applied having no significant effect. The correspondance of major features with NIFS $K$ are clearly evident, particularly `H2' with `G2' and `L4' with `G3'.

\begin{figure*}
\plotonewide{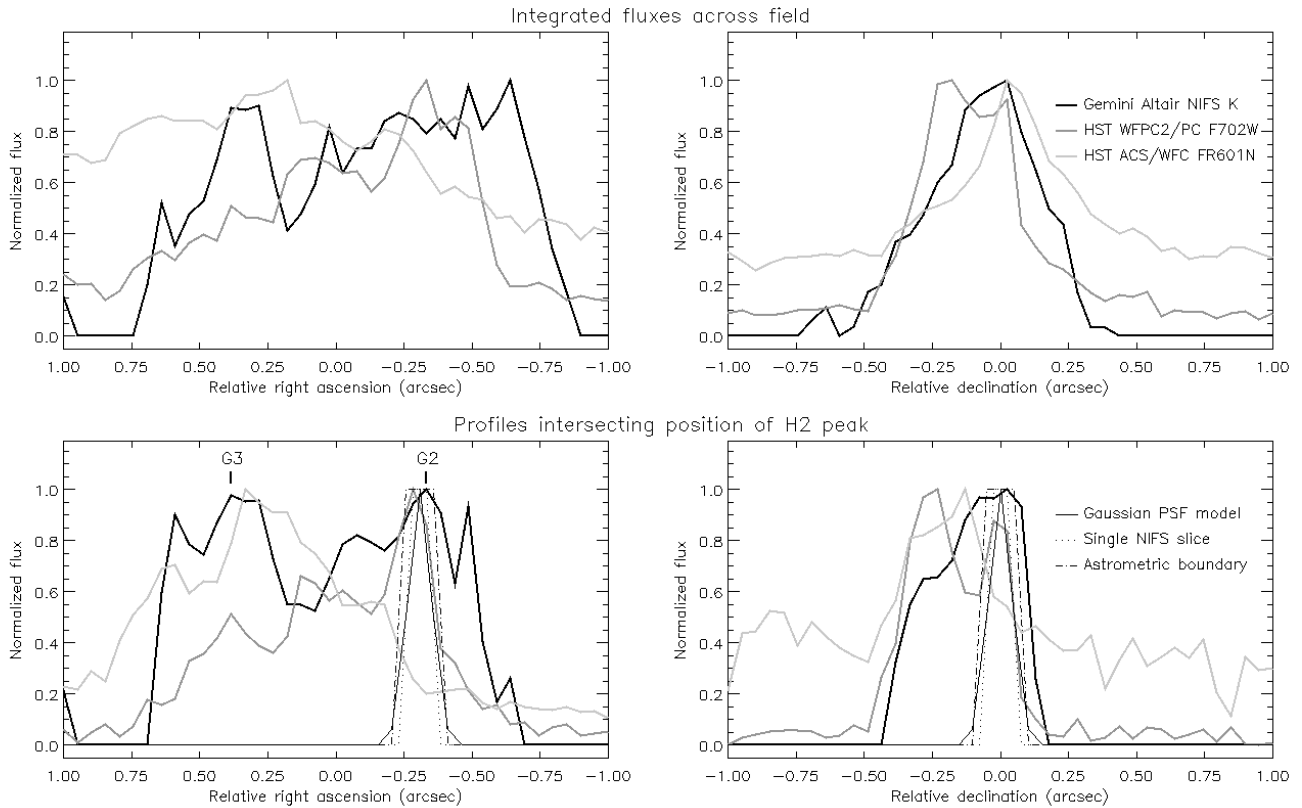}
\caption{Plots of the normalized flux integrated along right ascension and declination (top) and, similarily, across 0\farcs15-wide slices that intersect the position of H2 in the HST WFPC2/PC F702W image (bottom). The same for a Gaussian PSF of width 0\farcs10 and one representative of the nominal, single NIFS slice are overplotted. To within instrumental limits the positions of H2 and G2 correspond.}
\label{plot_profiles}
\end{figure*}

A lack of isolated point sources in the small field means that there is no easy way to provide an absolute registration to the $K$ image at sub-pixel accuracy, although the compactness and peak brightnesses of the structures in that and the HST images do provide an internal consistency check. This is illustrated in Figure~\ref{plot_profiles}: profiles are shown both as integrals across the field and as north-south/east-west slices of width 0\farcs15 (3 pixels) that intersect at the peak of H2 (thick curves). This is centred at ${\rm R.A}=$~06h50m52.167s ${\rm Dec.}=$~+41$^\circ$30\arcmin31\farcs19. That position is indiscernable from the peak of G2, within its resolution and pointing accuracy. A boundary (dot-dashed outline) in Figure~\ref{plot_profiles} is meant to indicate what might be due to misregistration to HST of the Gemini image; a combination of positional errors and the outside limit of spatial resolution: $\sqrt{(0\farcs043)^2+(0\farcs146)^2}\approx0\farcs15$. Also overplotted are the width of a single NIFS slice (dotted outline) and the estimated PSF (thin black curve). Compared to any of those limits, some structures appear slightly ``peakier", e.g. the western edge of G2. This suggests that 0\farcs10 is a good estimate of the combined spatial resolution and positional accuracy, presuming structures sample a common optical/NIR spectrum although emission is from marginally resolved knots.

\subsection{Spectra}\label{spectra}

\begin{figure}
\plotone{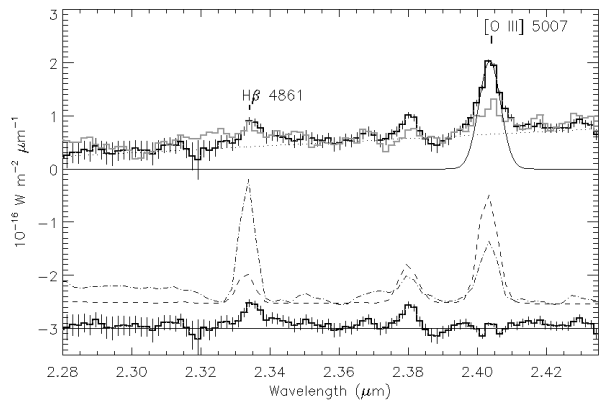}
\caption{A spectrum of 4C+41.17 from the NIFS datacube, integrated over all pixels (black), and just the 9 pixels centered-on G2 (dark grey); Gaussian fit to [O {\small III}]$\lambda$5007 indicated by the solid curve, continuum by a dotted line. Template spectra (dashed) are discussed in the text; results after subtraction below.}
\label{plot_spectrum}
\end{figure}

Spectra were then extracted at each spatial pixel in the datacube. The result of integrating those over high-$S/N$ pixels in the centre of the datacube is shown in the top panel of Figure~\ref{plot_spectrum}; error-bars are formal limits for Poisson noise only. Significant continuum emission is seen, and the ratio of [O {\small III}]$\lambda$5007 to H$\beta$ is about 2.2 (the other line in the doublet is weak) typical of HzRGs. From the composite spectrum of \cite{Eales1993} this ratio might be expected to be as high as 3, for example with Ly$\alpha$/H$\beta$ $\approx 6$ and [O {\small III}]$\lambda$5007/Ly$\alpha$ $\approx 0.6$. 

Nebular emission and star-formation history among the knots probably varies though, as could dust extinction. To illustrate this, two comparison spectral templates from the Kinney-Calzetti spectral atlas of galaxies \citep{Calzetti1994, Kinney1996} are plotted with arbitrary normalization, starbursts with the widest-available range of colours with dust extinction: one with $E(B-V) < 0.1$ (dashed) and the other with $0.61 < E(B-V) < 0.70$ (dot-dashed). A spectrum accumulated from just the $3\times3$ pixel box centred at H2 (G2) is overplotted as a dark-grey curve. The thin solid curve indicates a fit of a Gaussian to the [O {\small III}]$\lambda$5007 line, the convolution of instrumental resolution and line-width; the residual after subtraction from the spectrum is shown below. Starbursts provide a good match to the spectra of bright knots near the galaxy core.

\subsection{Kinematics, Line Strengths, and Continuum}\label{kinematics}

\begin{figure*}
\plotonewide{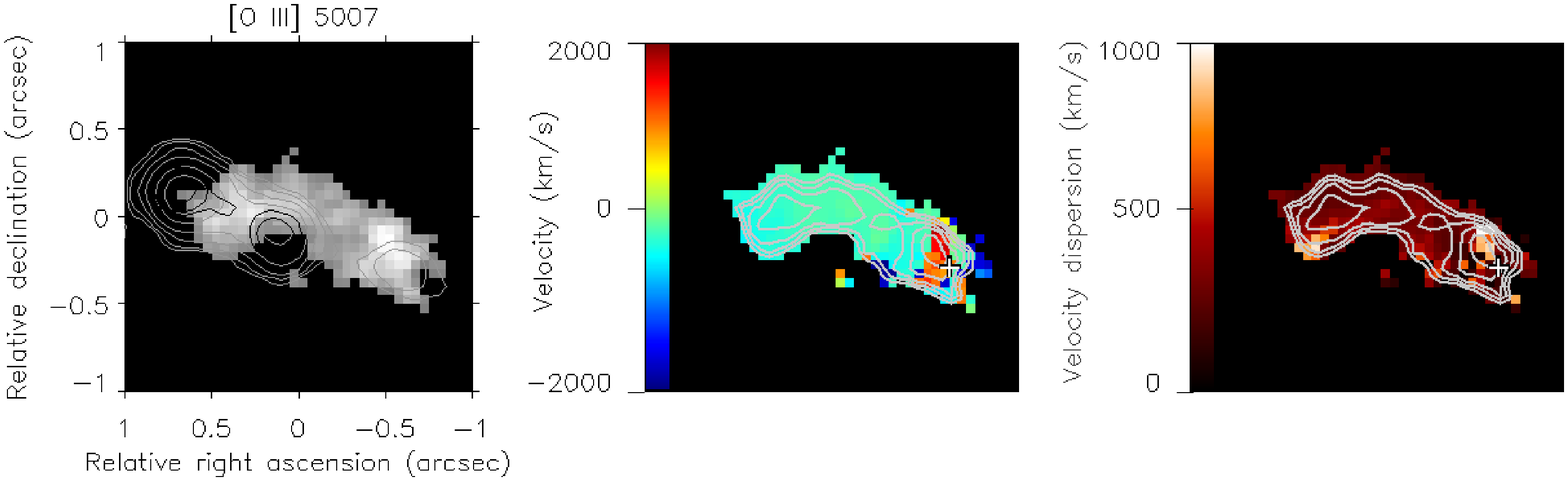}\\
\plotonewide{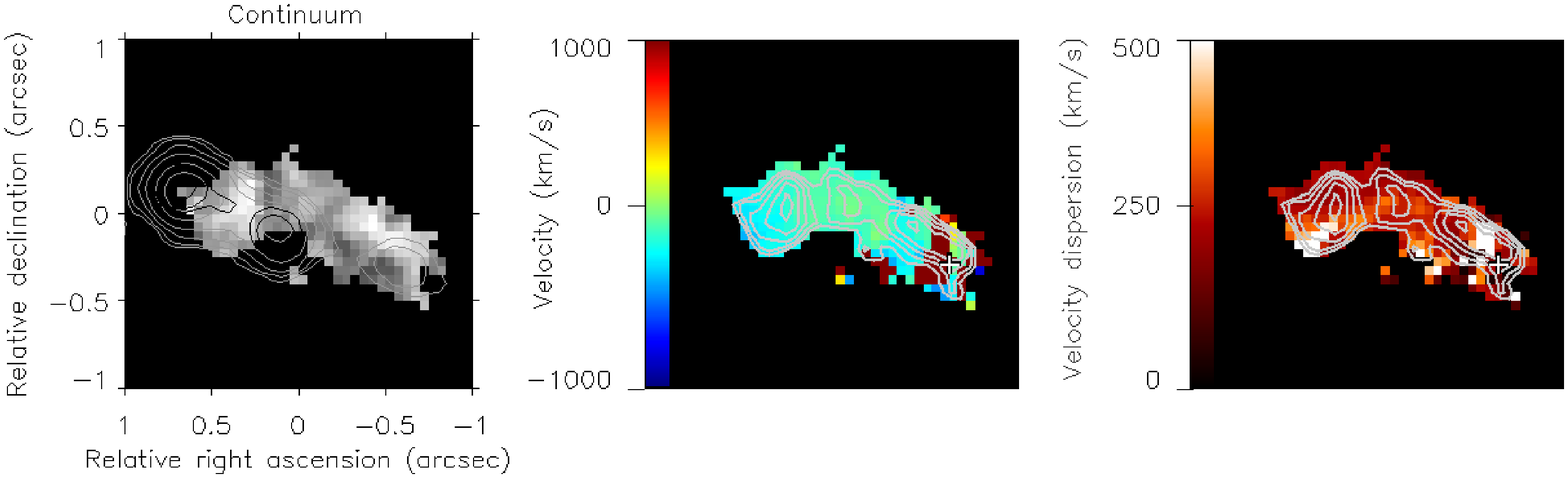}\\
\plotonenarrow{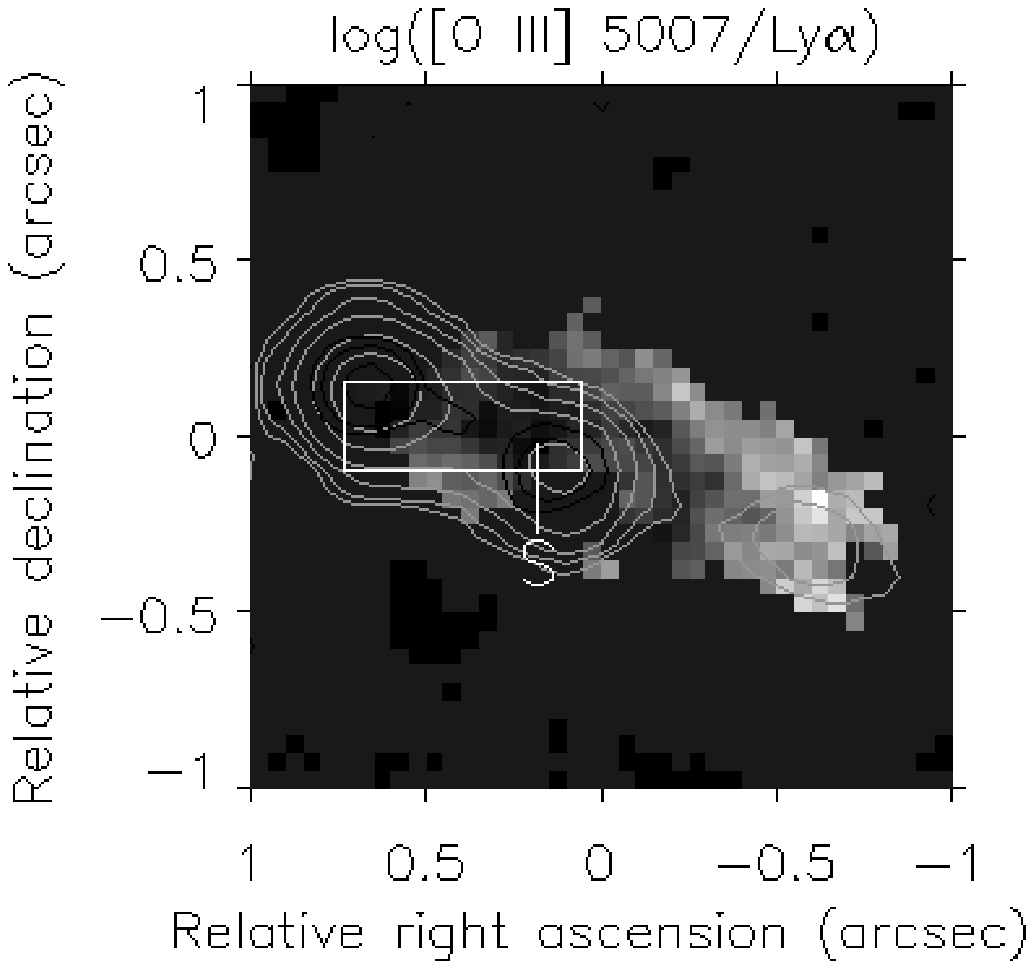}\\
\caption{Gemini NIFS $K$ band separated into [O {\small III}]$\lambda$5007 emission (top left panel) and continuum (middle left panel); 8 GHz (grey) and 15 GHz (black) radio maps are overplotted. To the right are velocity maps of [O {\small III}]$\lambda$5007 emission overplotted with corresponding photometric contours (in 10\% increments above 50\% peak brightness) for [O {\small III}]$\lambda$5007 and continuum; the nucleus is marked with a cross. Note how the brightest continuum emission lies directly between, and in line with, radio components B2 an B3; along a perpendicular ridge of elevated velocity dispersion. Bottom: map of [O {\small III}]$\lambda$5007 to Ly$\alpha$ line strength; box inscribes pixels between positions of B2 and B3.}
\label{plot_velocities}
\end{figure*}

Kinematic information and line strengths were recovered at each pixel by a Gaussian fit to the [O {\small III}]$\lambda$5007 line position and width. Continuum flux was estimated by subtracting this fit from the spectrum; the weaker line in the doublet and H$\beta$ were blocked out. The results are mapped in Figure~\ref{plot_velocities}, separated into two parts. The [O {\small III}]$\lambda$5007 emission is shown on the left of the top row, and immediately below is remaining continuum. Overplotted are the radio maps at 8.3 GHz and 15 GHz. To the right are velocities and velocity dispersions (after subtracting the spectral resolution in quadrature) overplotted with flux contours, shown in two stretches: along the top row for the full range of data up to $2000~{\rm km}~{\rm s}^{-1}$, and restricted by a factor of two below. 

Both broad (${\rm FWHM}\approx 1000~{\rm km}~{\rm s}^{-1}$) and narrow ($\approx500~{\rm km}~{\rm s}^{-1}$) components evident in long-slit Keck sprectroscopy \citep{Reuland2007} are seen here. The region of highest turbulence ($>1500~{\rm km}~{\rm s}^{-1}$) occurs very near the nucleus (marked with a cross in Figure~\ref{plot_velocities}) with most of the galaxy exhibiting fairly low velocities relative to systemic. Another thing to note is that the distributions of [O {\small III}]$\lambda$5007 emission and $K$-band continuum differ slightly. For example, near the nucleus, the continuum light centred at G2 is more elongated, arcing towards the east. In [O {\small III}]$\lambda$5007 emission, this shape continues to the west around B2, tracing towards G3, in good agreement with the elliptical shock described in \cite{Bicknell2000}. In the region of G3, between B2 and B3, the continuum emission seems flattened, stretching south and north, akin to a thin sheet or shell, perpendicular to the radio axis.  That structure is also mirrored in the kinematics: somewhat higher, turbulent velocities follow a narrow ridge of brighter continuum flux. This is especially evident in the right-hand panel of the middle row, highlighting velocity dispersions near the median of $241~{\rm km}~{\rm s}^{-1}$, those over $500~{\rm km}~{\rm s}^{-1}$ appear ``saturated" in this stretch.

Finally, the ratio of [O {\small III}]$\lambda$5007 to Ly$\alpha$ emission was estimated at each pixel by dividing the [O {\small III}]$\lambda$5007 image by ACS/WFC FR601N, which is normalized to a Ly$\alpha$ flux of $2.2\times10^{-15}~{\rm erg}~{\rm s}^{-1}~{\rm cm}^{-2}$ and equivalent line width of 1300\AA. The result is shown in the bottom panel of Figure~\ref{plot_velocities}, of special interest is the region directly along a narrow ``corridor" from B2 to B3, discussed more in the next section. This is surrounded by an envelope with the appearance of an edge-brightened cocoon, where higher [O {\small III}]$\lambda$5007/Ly$\alpha$ occurs. This is strongly indicative of star-formation regions (values over 3 appear white in this stretch), particularly near the nucleus and G2, which have spectra similar to a starburst.

\section{Discussion and Summary}\label{discussion}

The new Gemini restframe optical images show a close resemblence to UV emission seen with HST. The main structures near the core, G2 and G3, are clumpy and have spectra consistent with local starburst galaxies, and from comparison to VLA maps, it is evident that the radio jet is involved in shaping this morphology. Furthermore, the kinematics and emission-line ratios along the axis of the jet between B2 and B3 set it apart as distinct from the rest of the galaxy, which would not have been resolved with seeing-limited spectroscopy.

\begin{figure*}
\plotonewide{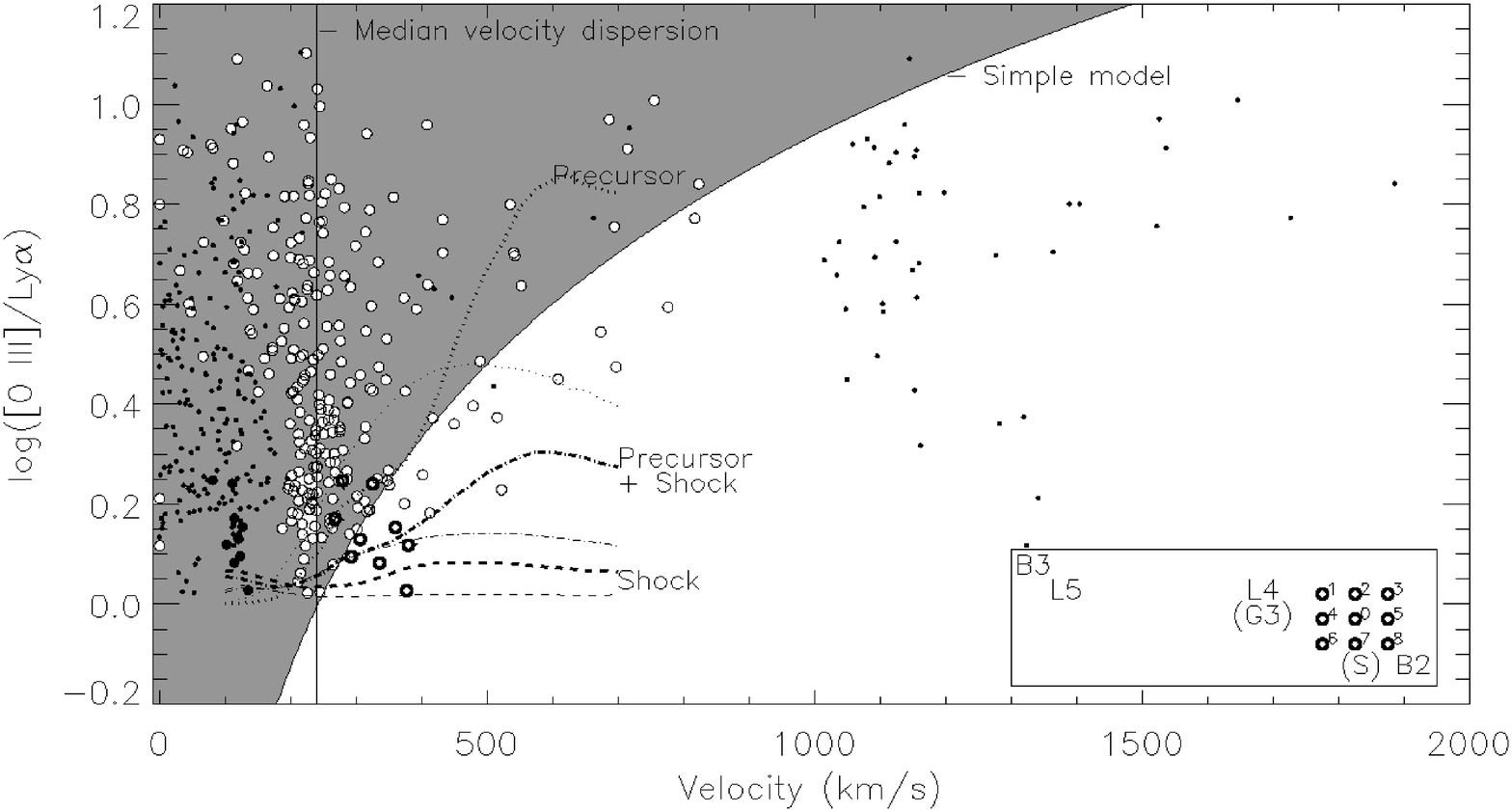}\\
\plotonewide{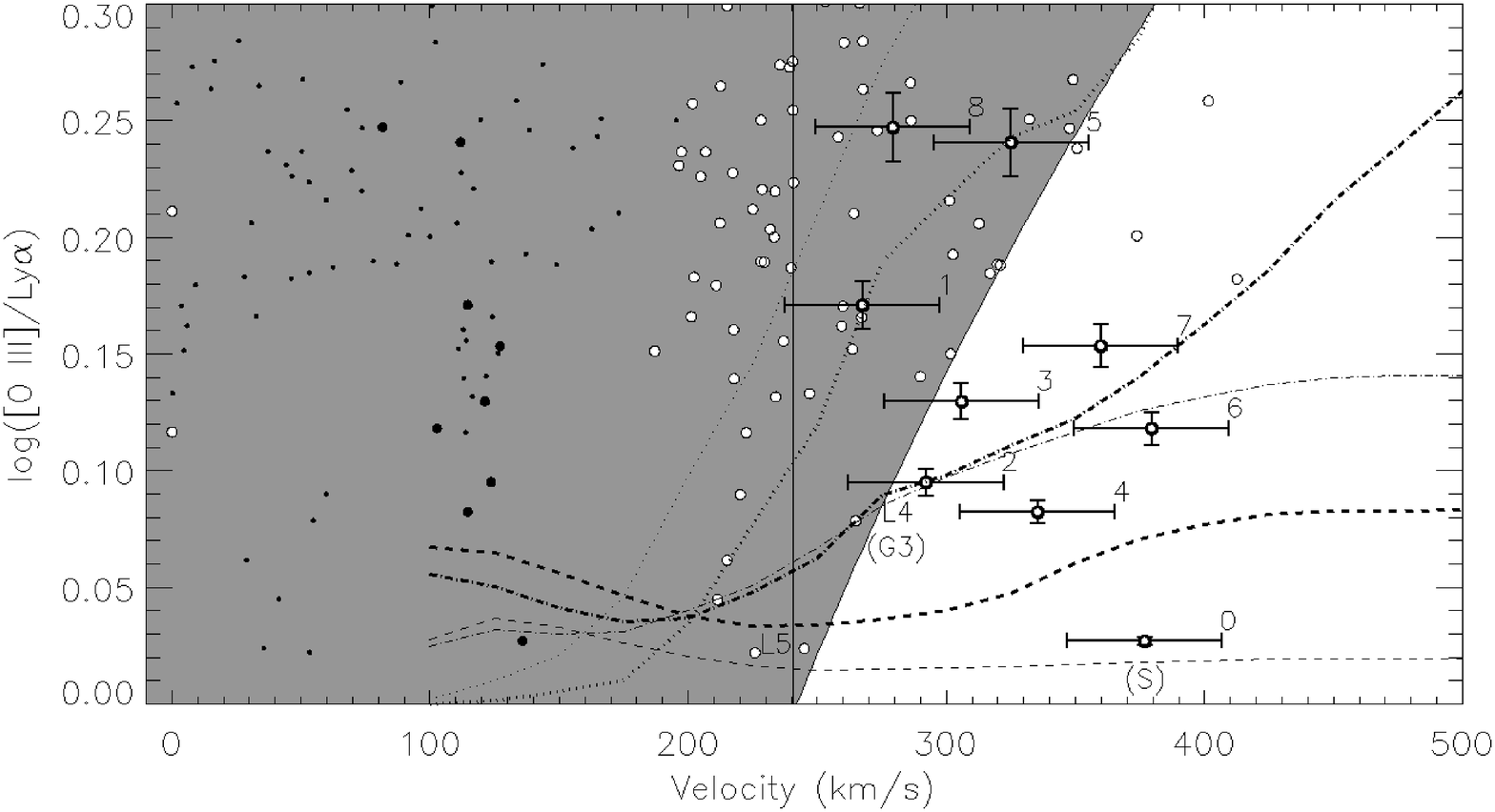}
\caption{Ratios of [O {\small III}]$\lambda$5007 to Ly$\alpha$ emission, plotted relative to absolute velocity (black dots) and velocity dispersion (open circles) at each spatial pixel. A scaled box in the upper panel indicates the ``corridor" between the pixels inscribed by those corresponding to B2 and B3; the lower panel is restricted to results within this box (also outlined in Figure~\ref{plot_velocities}). Highlighted are the pixel with lowest ratio of [O {\small III}]$\lambda$5007 to Ly$\alpha$ and its eight nearest neighbours, shown with larger symbols in velocity and error bars in velocity dispersion, numbered 1 through 8. Model curves are discussed in Section~\ref{discussion}.}
\label{plot_velocity_correlation}
\end{figure*}

To help elaborate on the special properties of the jet corridor relative to the rest of the galaxy, Figure~\ref{plot_velocity_correlation} plots the ratio of [O {\small III}]$\lambda$5007 to Ly$\alpha$ against velocity in each pixel; the absolute value of velocity is indicated as small black dots and velocity dispersion as larger open circles. Although long-slit spectroscopy may have led to overestimated shock velocities due to ``contamination" from outside, spatial and spectral resolution here is sufficient to discern deviations from the median velocity dispersion within the jet corridor. The small inset box indicates relative positions: a convenient choice is to have B2 and B3 at opposite corners, as that way each horizontal row is then one pixel high (and each vertical column a pixel wide). Labels within the box are also approximately the size of a single pixel (0\farcs05) in the dataset; in cases of overlap, labels appear in brackets immediately below the pixel in the HST image to which they have been identified. Misregistration of the radio maps to these identifications would be relative shifts less than a pixel. The rectangle is 0\farcs65 long and 0\farcs25 wide (projected area of 4.7 kpc by 1.8 kpc) and the bottom panel of Figure~\ref{plot_velocity_correlation} is confined to that zoomed-in spatial region, i.e. all pixels inside fall within these restricted axes ranges.

Note in particular that turbulence within the jet corridor is generally less than outside apart from a few pixels. Remarkably, these are from the same compact area and share a similar low velocity relative to systemic for a range of line indices. The location where minimum ratio of [O {\small III}]$\lambda$5007 to Ly$\alpha$ is reached, yet at highest velocity dispersion, is at the pixel 0\farcs10 east and 0\farcs10 north of B2, a projected distance of only 1.0 kpc - marked by `S' in Figure~\ref{plot_velocity_correlation}. This location is a peak in the [O {\small III}]$\lambda$5007 emission, so although dust may be present, a dust gradient (strong obscuration of Ly$\alpha$) does not seem to be a likely explanation. The pixel S, and the eight nearest pixels surrounding it (labelled 1 through 8) are just resolved from B2, based on the PSF and astrometric analysis of Section~\ref{astrometry}. Error bars are similar for other measurements, but omitted for clarity. To the northeast of S is L4, which is coincident with G3 and the beginning of a faint tail of 15 GHz emission stretching back towards B2 from B3, the radio lobe (most easily seen in the synthetic $K$-band image in the lower-right panel of Figure~\ref{plot_images}). This detailed correspondance - the tip of a coherent structure in kinematics and line ratio directly along the radio jet - points to S being the location of the shock associated with the jet head itself.

In the model of shock emission described in \cite{Bicknell2000}, the energy flux $F$ deposited by jet momentum grows by the cross-sectional area $A$ of interaction with the (possibly jittering) jet, that is $$F\propto A~n_{\rm H}~v_{\rm sh}^2. \eqno(1)$$ The efficiencies of line emission for shocked gas (for both relativistic and non-relativistic jet cases) are given by shock coefficients. These have been calculated by \cite{Dopita1995, Dopita1996} at low shock velocities ($<500~{\rm km}~{\rm s}^{-1}$) and magnetic field parameterization $B/n^{1/2}\leq4~\mu{\rm G}~{\rm cm}^{-3/2}$. Shown in Figure~\ref{plot_velocity_correlation} is the expected range of line ratio consistent with \cite{Bicknell2000} for the shock coefficient $\alpha$(C~{\small IV}). That is roughly linear in a log-log plot with velocity (their Figure 4), with a slope of 1.52, although with an upturn for lower metallicity and higher shock velocity. The observed C {\small IV}/Ly$\alpha$ ratio is 0.091 integrated over the galaxy \citep{Chambers1990} but values within the corridor are unknown. A plausible corresponding lower limit is simply set by $\log$([O~{\small III}]/Ly$\alpha$)$=\log{v_{\rm sh}^{1.52}}$ as indicated by the shaded region in Figure~\ref{plot_velocity_correlation}. It may be coincidence that the bulk of stellar light (at G3) also falls on this curve, but it does suggest that [O~{\small III}] emission and kinematics at both L4 and L5 can be consistent with the previous interpretation of Ly$\alpha$ emission in the jet corridor.

Some refinement to this picture might be gleaned from comparison of higher resolution models to the data. Newer line-ratio shock model grids \citep{Allen2008} at finer mesh than the earlier Dopita \& Sutherland ones can also be compared. These are fully radiative shock models calculated with the MAPPINGS III shock and photo-ionization code, which extend to lower metallicities and higher shock velocities. In Figure~\ref{plot_velocity_correlation} a range of metallicity parameter space is plotted: a model with nominal values of solar abundance and maintaining $B/n^{1/2}=3.23~\mu{\rm G}~{\rm cm}^{-3/2}$ is shown in phases of preshock (dotted), shock (dashed) and the combination of the two: preshock plus shock (dot-dashed). Curves for higher $B/n^{1/2}$ (not shown) cannot match the low line ratio at S, although this involves some degeneracy between field strength and gas density. Plausibly, a range of values occurs within the galaxy core. Lower-metalicity models following the (LMC-like) range in \cite{Dopita2005} are indicated by the thin curves. Note that even for lower metallicity, the pixel at S still falls within the shock region, while those nearby may possibly be in a mixed state; those further away always belong to the precursor region.

Notably, within the shell structure associated with L4, the brightest continuum emission at G3 appears ``downstream" along the jet from the shock front. A starburst here is expected, but it is also reminiscent of the ``lobe" of stars as interpreted by \cite{Steinbring2011} in previous ALTAIR NIFS spectrocopy of 3C 230, another extremely massive HzRG. Whether in 4C+41.17 star-formation takes place in situ in the shocked gas, or that outflow related to the jet may also be underway is unclear, but what is interesting is that the star-formation region it represents is compact and of relatively low mass. And evidently, there could be many such starbursting knots near the core unresolved in these data. Just a simple order of magnitude estimate of the mass in a single pixel, if the velocity dispersion is assumed to be comparable to the escape velocity, would be $$M_{\rm pixel} = {{5 R \sigma^2}\over{G}}\approx 2\times10^{10} M_\odot. \eqno(2)$$  A star-formation region the size of G3 (a few pixels) could encompass a mass of $\sim10^{11} M_\odot$, roughly consistent with the young stellar population discussed in \cite{Rocca-Volmerange2013}.

In summary, 4C+41.17 has been observed with Gemini ALTAIR NIFS spectroscopy, the highest redshift RG which has been studied in this way. A careful choice of IFU orientation and dither positions provided the best-possible AO spatial resolution, allowing detailed correspondance with previous HST WFPC2 and ACS optical images and VLA radio maps. Taken together, the kinematics, line-emission and continuum support the view that shock-induced starbursts are underway in this galaxy, and confirms a previous interpretation of a bow-shaped structure seen in Ly$\alpha$ emission being the leading edge of a shock front. This highlights for 4C+41.17 the intricate connection between jet-induced shocks, star-formation, and galaxy morphology among massive HzRGs. Dissentangling magnetic field strength from density variation in shock models would benefit from deeper spectra, especially by allowing more direct tests using H$\beta$ line diagnostics. Future studies in the coming era of 30-m class telescopes could resolve these knots at the level of $10^{9} M_\odot$, which requires resolution roughly twice as sharp as provided here: PSF ${\rm FWHM}\approx0\farcs05$. One challenge will be accurate astrometry, which in the current study was limited by the positional accuracy of the telescope plus AO and IFU, together with the control of the PSF. Improvement of that might come from real-time monitoring of AO performance via WFS telemetry, but also possibly by resolving individual compact star-forming clumps.

\acknowledgements

I thank the staff of Gemini Observatory, particularly observer Andy Stephens, and Richard McDermid for his assistance in getting the blind offsets just right. Thoughtful comments from an anonymous referee helped improve the original manuscript. This research used the facilities of the Canadian Astronomy Data Centre operated by the National Research Council of Canada with the support of the Canadian Space Agency. This work is based in part on observations obtained at the Gemini Observatory, which is operated by the Association of Universities for Research in Astronomy, Inc., under a cooperative agreement with the NSF on behalf of the Gemini partnership: the National Science Foundation (United States), the Science and Technology Facilities Council (United Kingdom), the National Research Council (Canada), CONICYT (Chile), the Australian Research Council (Australia), Minist\'{e}rio da Ci\^{e}ncia e Tecnologia (Brazil) and Ministerio de Ciencia, Tecnolog\'{\i}a e Innovaci\'{o}n Productiva (Argentina).

\end{document}